\documentclass[a4paper,12pt]{article}
\usepackage{amsmath,amsfonts,amssymb,epsfig,subfigure, color}

\newcommand{\be}{\begin{equation}}
\newcommand{\en}{\end{equation}}

\newcommand{\demi}{\textstyle{\frac{1}{2}}}
\newcommand{\troismi}{\textstyle{\frac{3}{2}}}
\renewcommand{\vec}[1]{\boldsymbol{#1}}

\begin{document}
\numberwithin{equation}{section}

\title{Scalar evolution equations for shear waves \\ in incompressible solids: \\ A simple derivation of the Z, ZK, KZK, \\and KP equations}


\author{Michel Destrade$^{1}$, Alain Goriely$^2$ \& Giuseppe Saccomandi$^3$\\[12pt]
$^1$School of Mathematics, Statistics \& Applied Mathematics,\\ 
National University of Ireland Galway,\\ University Road, Galway, \\Republic of Ireland;\\[12pt]
$^2$OCCAM,  Mathematical Institute,\\ University of Oxford, OX1 3LB, UK;\\[12pt]
$^3$Dipartimento di Ingegneria Industriale, \\ Universit\'a degli Studi di Perugia, 06125 Perugia, Italy.}

\date{}
\maketitle



\begin{abstract}

We study the propagation of two-dimensional finite-amplitude shear waves in a nonlinear pre-strained incompressible solid, and derive several asymptotic amplitude equations in a simple, consistent, and rigorous manner. 
The scalar Zabolotskaya (Z) equation is shown to be the asymptotic limit of the equations of motion for all elastic generalized neo-Hookean solids (with strain energy depending only on the first principal invariant of Cauchy-Green strain). 
However, we show that the Z equation cannot be a scalar equation for the propagation of two-dimensional shear waves in general elastic materials (with strain energy depending on the first and second principal invariants of strain). 
Then we introduce dispersive and dissipative terms to deduce the scalar Kadomtsev-Petviashvili (KP),  Zabolotskaya-Khokhlov (ZK) and Kho\-khlov-Za\-bo\-lots\-ka\-ya-Kuz\-net\-sov (KZK) equations of incompressible solid mechanics.

\end{abstract}


\newpage


\section{Introduction}


The Zabolotskaya (Z) equation describes the propagation of finite-amplitude, two-dimensional, linearly-polarized, shear waves in nonlinear solids. It can be presented in the following form (due to Cramer and Andrews (2003)):
\be \label{Z}
\left( U_\tau + \beta U^2 U_\chi \right)_\chi + \demi U_{\eta \eta} = 0,
\en
where $U= U(\chi, \eta, \tau)$ is a measure of the amplitude of the wave, $\chi$ is a spatial variable measured in a frame moving with the wave, $\eta$ is a spatial variable transverse to the wave propagation, $\tau$ is a temporal variable, and $\beta$ characterizes the material properties of the solid. 
Zabolotskaya (1986) derived it via a perturbation scheme from the general equations of elastodynamics. 
It is a counterpart to the Zabolotskaya-Khokhlov (ZK) equation, itself describing the propagation of weakly nonlinear, weakly diffracting, two-dimensional sound beams (Zabolotskaya and Khokhlov, 1969):
\be \label{ZK}
\left( U_\tau + \beta U U_\chi \right)_\chi + \demi U_{\eta \eta} = 0.
\en
(Note that this form is also due to Cramer and Andrews (2003), and that the nonlinear parameters $\beta$ are not the same in each equation).
The history of the ZK equation and its variants has recently been retraced by Rudenko (2010).

There are two major differences between the Z and ZK equations. 
One is obviously the cubic term in \eqref{Z} versus the quadratic term in \eqref{ZK}.
The other is that the ZK equation governs the motion of the \emph{longitudinal} displacement and is a \emph{scalar equation}, while the Z equation governs the propagation of plane-polarized \emph{shear waves} and is thus a \emph{vector equation} a priori.
Nonetheless, recent papers (Enflo et al., 2006; Wochner et al., 2008; Pinton et al., 2010) have investigated the propagation of linearly-polarized nonlinear shear waves in incompressible solids. 

Here we aim at finding what restrictions must be imposed, if at all, on the form of the strain energy $W$ (say), in order to obtain a \emph{scalar Z equation}. 
We also include the effects of pre-strain, similar to Cramer and Andrews (2003) in the compressible case. 
We point out that another problem of interest related to these model equations concerns the mathematical validation of the various approximations. 
This problem has been solved in the framework of the Navier-Stokes and Euler system by Rozanova-Pierrat (2009).
Moreover, because we are mainly interested in the understanding the physical underpinnings at the basement of various modelling assumptions, we adopt here a direct method for the derivation of the model equations, not a general reductive perturbation scheme such as the one proposed by Taniuti (1990) for example. 

In \S\ref{the-z-equation} we establish that a scalar Z equation can indeed be generated for the whole class of the so-called \emph{generalized neo-Hookean solids}, for which the strain-energy density does not depend on the second principal invariant of the Cauchy-Green strain tensor. 
We then try to generalize the study to the case of a generic \emph{fourth-order incompressible solid}, for which $W$ is expanded up to the fourth order in invariants of the Green strain tensor. 
We show in \S\ref{the-z-equation}\ref{general-4th-order} that there, the Z equation is necessarily a vector equation that is, only plane-polarized---not linearly-polarized--- two-dimensional shear waves are possible.

When dissipative and dispersive effects are taken into account for the propagation of longitudinal waves, the ZK equation is generalized to
\be \label{KZKP}
\left( U_\tau + \beta U U_\chi + \beta' U U_{\tau \tau} + \beta'' U_{\tau \tau \tau}  \right)_\chi + \demi U_{\eta \eta} = 0,
\en
where $\beta'$ and $\beta''$ are other material constants.
This scalar equation covers the purely dissipative (at $\beta''=0$) Kho\-khlov-Za\-bo\-lots\-ka\-ya-Kuz\-net\-sov (KZK) equation (Kuznetsov, 1971) and the purely dispersive Ka\-dom\-tsev-Pet\-vi\-ash\-vi\-li (KP) equation (Kadomtsev and Petviashvili, 1970).
In \S\ref{dissip-disper} we consider dissipative and dispersive effects in generalized neo-Hookean solids. 
We derive the counterpart to \eqref{KZKP} for shear waves, as
\be \label{KZKPshear}
\left( U_\tau + \beta U^2 U_\chi + \beta' U U_{\tau \tau} + \beta'' U_{\tau \tau \tau}  \right)_\chi + \demi U_{\eta \eta} = 0,
\en
a scalar equation which covers the Z equation, the modified Kho\-khlov-Za\-bo\-lots\-ka\-ya-Kuz\-net\-sov (mKZK) equation and the modified Ka\-dom\-tsev-Pet\-vi\-ash\-vi\-li (mKP) equation.

The aim of this paper is to provide a simple derivation of these scalar equations for incompressible materials. This quest is largely motivated by the pressing need to model acoustic wave propagation in soft tissues and gels, where the combined effects of incompressibility, pre-strain, viscosity, and dispersion cannot be ignored. This is a topic of fundamental relevance to ultrasound-based techniques of medical imaging such as elastography.


\section{Basic Equations}


Consider a body of incompressible isotropic elastic material, initially at rest in a reference configuration $\mathcal{B}_0$, say.
For the study of anti-plane shear motions, we choose a system of Cartesian coordinates $(X_1, X_2, X_3)$ to denote the position of a particle in the body in $\mathcal{B}_0$.
We then call ($x_1, x_2, x_3$) the Cartesian coordinates, aligned
with ($X_1, X_2, X_3$), corresponding to the current position $\vec{x}$ in the deformed configuration $\mathcal{B}$, say.
We study the anti-plane motion 
\be \label{shear_deformation}
x_1 = \lambda^{-1/2} X_1, \qquad 
x_2 = \lambda^{-1/2} X_2, \qquad 
x_3 = \lambda X_3 + u(X_1, X_2, t),
\en
where $\lambda$ is a positive constant---the axial pre-stretch---and $u$ is a  three-times continuously differentiable function---the shear motion.

We call $\vec{F} \equiv \partial \vec{x}/\partial \vec{X}$ the associated deformation gradient, and $\vec{B}= \vec{F} \vec{F}^t$ the left Cauchy-Green
strain tensor. 
The first principal isotropic strain invariant,
$I_1 \equiv \text{tr}\vec{B}$, is  here 
\be
I_1 = \lambda^2+2 \lambda^{-1} + u_1^2 + u_2^2,
\en
and the second principal isotropic strain invariant,
$I_2 \equiv \demi [I_1^2 - \text{tr}(\vec{B}^2)]$, is equal to
\be
I_2 = \lambda^{-2}+2 \lambda + \lambda^{-1}(u_1^2 + u_2^2),
\en
where $u_1 \equiv \partial u/\partial X_1$, $u_2 \equiv \partial u/\partial X_2$.
In all generality, the strain energy density $W$ is a function of $I_1$ and $I_2$ only, and the Cauchy stress tensor $\vec{\sigma}$ is derived as (see e.g. Ogden, 1984),
\be \label{general}
\vec{\sigma} =  - p \vec{I} + 2W_1 \vec{B} - 2 W_2 \vec{B}^{-1},
\en
where $W_1 \equiv \partial W / \partial I_1$, $W_2 \equiv \partial W / \partial I_2$, and $p$ is a Lagrange multiplier introduced by the constraint
of incompressibility (which imposes that $\det \vec{F}=1$ at all times).

The equations of motion, in absence of body forces, are 
\be
\text{div }\vec{\sigma} =\rho \partial^2\vec{x}/\partial t^2,
\en
where $\rho$ is the mass density, which remains constant due to the incompressibility constraint.  
Here, the equations of motion reduce to the following system of partial differential equations (see Knowles (1976) for the static case),
\begin{eqnarray}\label{equation}
&& q_j + 2\left(W_2 \ u_j u_i\right)_i=0, \qquad j=1,2
\\
&&2[ (W_1 + \lambda^{-1} W_2) \ u_i]_i=\rho \partial^2 u/\partial t^2,
\end{eqnarray}
where summation over the repeated index  $ i=1,2$ is assumed. 
The auxiliary quantity $q$ is given by
\be
 q =\lambda p(X_1, X_2)-2W_1-2(\lambda^2+\lambda^{-1}+ u_1^2 + u_2^2)W_2.
 \en


\section{The Z equation}
\label{the-z-equation}

\subsection{Results for generalized neo-Hookean materials}


The system \eqref{equation} consists of three partial differential equations for the two unknown functions $p$ and $u$.  Therefore the system is overdetermined and it is not possible, in general, to find a general solution for arbitrary choices of the strain energy $W$. Progress is possible for some special forms of the strain energy density, in particular for the Mooney-Rivlin form, which is linear in $I_1$ and $I_2$ (Adkins, 1954), but only for \emph{static} solutions, see Knowles (1976) for details. The dynamic situation for antiplane \emph{motions} is much more complex, as discussed by Hayes and Saccomandi (2004) in the Mooney-Rivlin case. 
 
However, if we consider materials for which $W=W(I_1-3)$ only---the so-called \emph{generalized neo-Hookean materials}---, then the system \eqref{equation} reduces to a well-determined system
\be \label{eqn}
q_1=q_{2}=0, \qquad 
2(W' u_i)_i= \rho \partial^2 u / \partial t^2,
\en
where $W'$ is the derivative of $W$ with respect to its argument $(I_{1}-3)$.


\subsubsection{Derivation of the Z equation}


To derive the Z equation, we assume that $W$ is at least of class $\mathcal{C}^{3}$. We   assume that the primary propagation direction is in the $X_1$-direction and  introduce the following new displacement and new variables
\be \label{changes2}
v = \epsilon^{-1} u, \qquad
\chi = \epsilon^2 {X}_1, \qquad 
\tau = t \pm \alpha {X}_1, \qquad 
\eta = \epsilon \sqrt{\alpha} {X}_2, 
\en
where $\epsilon$ is a small, non-dimensional parameter, giving a measure of the motion amplitude, the plus or minus sign corresponds to left or right running waves, and $\alpha$ is a parameter to be adjusted to cancel all terms to order $\epsilon$
(Note that the scalings used here mirror those of Zabolostkaya (1986), and give a consistent asymptotic expansion). 
To obtain the 
$Z \ equation$, we expand the equation \eqref{eqn}, that is
\begin{equation}\label{myeq}
2(W' u_{11}+W'_{1} u_{1}+W' u_{12}+W'_{2} u_{2})= \rho \partial^2 u / \partial t^2,
\end{equation}
to order $\epsilon^{3}$ (note again that the subscript $(\ )_{i}$ here denotes derivation w.r.t. $X_{i}$).

First, we compute the partial derivatives as follows,
\begin{align}
&  \dfrac{\partial}{\partial {X}_1} =\pm \alpha\dfrac{\partial}{\partial \tau}+ \epsilon^2 \dfrac{\partial}{\partial \chi}, \notag \\ 
&\dfrac{\partial}{\partial {X}_2} = \epsilon \sqrt{\alpha} \dfrac{\partial}{\partial \eta},\notag \\
& \dfrac{\partial^{2}}{\partial t^{2}} = \dfrac{\partial^{2}}{\partial \tau^{2}},\notag \\
& \dfrac{\partial^2}{\partial {X}_1^2} =\alpha^2 \dfrac{\partial^2}{\partial \tau^2}+ \pm 2 \epsilon^2 \alpha\dfrac{\partial^2}{\partial \chi \partial \tau} + \mathcal{O}\left(\epsilon^4\right),\notag \\
& \dfrac{\partial^2}{\partial {X}_2^2} =\epsilon^{2}\alpha \dfrac{\partial^{2}}{\partial \eta^{2}},
\end{align}
which leads to
\begin{align}
& u_{1}=\pm\epsilon \alpha v_{\tau}+\epsilon^{3} v_{\chi},\quad u_{2}=\epsilon^{2} \sqrt{\alpha} v_{\eta},\notag \\
& u_{11}=\epsilon \alpha^{2} v_{\tau\tau}\pm 2 \epsilon^{3} v_{\chi\tau}+\mathcal{O}\left(\epsilon^4\right),\quad u_{22}=\epsilon^{3} \alpha v_{\eta\eta}.
\end{align}

Second, we expand $W'$ in $\epsilon$. 
Since 
\begin{equation}
I_{1}-3=\lambda^{2}+2 \lambda^{-1}-3+\epsilon^{2} \alpha^{2}v_{\tau}^{2}+\mathcal{O}\left(\epsilon^4\right),
\end{equation}
we have 
\begin{equation}
W'=w^{(1)}+\epsilon^{2} w^{(2)}  \alpha^{2}v_{\tau}^{2} +\mathcal{O}\left(\epsilon^4\right)
\end{equation}
where $w^{(1)} = W'(\lambda^{2}+2 \lambda^{-1}-3)$ and $w^{(2)}=W''(\lambda^{2}+2 \lambda^{-1}-3)$.
Third, we use the expansion for $W'$ and $u_{i}$ to compute each term in~(\ref{myeq}) up to order $\epsilon^{3}$. That is
\be
W'_{1}=\pm \epsilon^{2} w^{(2)}\alpha^{3}\left(v_{\tau}^{2}\right)_{\tau}+\mathcal{O}\left(\epsilon^4\right),
\qquad
W'_{2}= \epsilon^{3} w^{(2)}\alpha^{5/2}\left(v_{\tau}^{2}\right)_{\eta}+\mathcal{O}\left(\epsilon^4\right).
\en
Equation~(\ref{myeq}) reads now
\begin{eqnarray}
\epsilon v_{\tau \tau}(2 w^{(1)}\alpha^{2} -\rho) + 2\epsilon^{3}\alpha\left[w^{(1)}(v_{\eta\eta} \pm 2 v_{\chi\tau}) + w^{(2)}\alpha^{3} (v^{3}_{\tau})_{\tau}\right] + \mathcal{O}\left(\epsilon^4\right)=0.
\end{eqnarray}
Terms to order $\epsilon$ disappear by choosing $\alpha^{2}={\rho/( 2 w^{(1)})}$, which gives to order $\epsilon^{3}$,
\begin{equation}
v_{\eta\eta}\pm 2 v_{\chi\tau}+{w^{(2)}\over w^{(1)}}\alpha^{3} (v^{3}_{\tau})_{\tau}=0.
\end{equation}

Finally, taking a derivative of this equation with respect to $\tau$ and using the substitution $U=v_{\tau}$ leads to the scalar Z equation:
\begin{equation}\label{equation4}
\left(\pm U_{\chi} + \beta U^{2} U_{\tau}\right)_{\tau}+ \tfrac{1}{2} U_{\eta\eta}=0,
\end{equation}
where $\beta=3 w^{(2)} \alpha^{5}/ \rho$.

For example consider the Yeoh strain energy described by 
\be \label{yeoh}
W = (\mu/2) \left[(I_1-3)+\gamma(I_1-3)^2/2\right],
\en
where $\mu>0$ is the initial shear modulus and $\gamma>0$ is a nonlinear elastic constant.
In this case we find
\begin{equation}
\alpha^{2}= {\rho\over \mu \left[1+\gamma(\lambda^2+2 \lambda^{-1}-3)\right]},
  \qquad \beta = \dfrac{3\mu}{2\rho}\gamma \alpha^5.
\end{equation}

Note that the apparent difference between \eqref{equation4} and the Z equation \eqref{Z} is simply due to the  choice of  independent variables.  
Here, we chose the same independent variables as Norris (1998),  where $\tau$ is a retarded/advanced time, $\chi$ is the spatial variable in the  propagation direction, and $\eta$ is the transverse spatial variable.

   
\subsubsection{Special solutions}

 
Following Cates and Crighton (1990) and Sion\'oid and Cates (1994), we perform the transformation 
\be \label{Cates}
U(\chi, \tau, \eta)=G(\chi, \zeta), \qquad  \text{where} \quad 
\zeta = \dfrac{1}{\beta}\left(\tau+\dfrac{\eta^2}{2\chi}\right),
\en   
which reduces the Z equation \eqref{equation4} to an equation which, once integrated in $\zeta$, yields
\be \label{equation4R}
G_{\chi} + G^2 G_\zeta +  \dfrac{1}{2 \chi}G = 0.
\en
The solution of this one-dimensional equation does not provide a general solution to the two-dimensional equation \eqref{equation4}; rather, it gives the solution to the particular Cauchy problems where initial data are given on the parabolic surfaces $\zeta= \text{const}$

A further  change of  function (from $G$ to $\hat G$) and of variables (from $\chi$ to $\hat \chi$), 
\be
\hat G(\hat \chi, \zeta) = \chi^{1/2} G \left(\chi, \zeta \right), \qquad 
\hat{\chi}=\ln \chi,
\en
reduces equation \eqref{equation4} to the autonomous inviscid generalized Burgers equation
\be
\hat{G}_{\hat{\chi}}+ \hat{G}^2\hat{G}_\zeta=0.
\en
The general integral of this equation is $\hat{G}=F\left(\zeta - \hat{G}^2\hat{\chi}\right)$, where $F$ is an arbitrary function. This allows us to recover the following class of exact implicit solution for the Z equation \eqref{equation4},
\be
U=\chi^{-1/2}F\left(\tau+\frac{\eta^2}{2\chi}-\beta U^2\chi \text e^\chi\right).
\en

It is quite remarkable that the transformation \eqref{Cates} should work for the Z equation (cubic nonlinearity), because it was initially intended for the KZK equation (quadratic nonlinearity). 
Through it, a huge class of exact solutions is generated for the Z equation. 

General methods to obtain exact solutions for such equations include reductions methods such as the one based on Lie group analysis and its generalizations. 
Those methods have been systematically applied to the ZK equations with quadratic nonlinearities (see for example Bruzon et al. (2009) and Tajiri (1995)), but it seems that there are very few results regarding the case of a cubic nonlinearity, of interest to incompressible elastic materials.
   

\subsection{Results for general incompressible solids}
\label{general-4th-order}


We return to an important point evoked in the Introduction: 
that the propagation of finite-amplitude transverse waves cannot be governed by a \emph{scalar} Z equation (i.e., linearly-polarized shear waves exist only in special solids, such as the generalized neo-Hookean materials). 

To show this, we take solids with the following Rivlin strain energy,
\be \label{MR2}
W = C_{10} (I_1-3) + C_{01}(I_2-3)+ C_{20}(I_1-3)^2,
\en
(where $C_{10}$, $C_{01}$, $C_{20}$ are constants), 
which covers the most general model of \emph{fourth-order incompressible elasticity} (Ogden, 1974), 
\be \label{incomp0}
W = \mu \; \text{tr}\left(\vec{E}^2\right)  + \frac{A}{3} \;\text{tr}\left(\vec{E}^3\right)   + D \; \left(\text{tr} (\vec{E}^2)\right)^2,
\en
with the identifications (Destrade et al., 2010):
\be \label{connections}
\mu = 2(C_{10} + C_{01}), \qquad 
A = -8(C_{10} + 2C_{01}), \qquad
D = 2(C_{10} + 3C_{01} + 2C_{20}).
\en
Here $\vec{E}=(\vec{F}^t\vec{F}-\vec{I})/2$ is the Green strain tensor and $\mu$, $A$, and $D$ are the second-, third-, and fourth-order elasticity constants, respectively.

With this strain energy, the first and second equations  of \eqref{equation} read
\begin{align} \label{two}
& \hat{q}_{\hat{X}_{1}} + 2 C_{01}\left[\left(\hat{u}^2_{\hat{X}_{1}}\right)_{\hat{X}_1} + \left(\hat{u}_{\hat{X}_{1}}\hat{u}_{\hat{X}_{2}}\right)_{\hat{X}_{2}} \right] = 0, \notag \\
& \hat{q}_{\hat{X}_{2}} + 2 C_{01}\left[\left(\hat{u}^2_{\hat{X}_{2}}\right)_{\hat{X}_2} + \left(\hat{u}_{\hat{X}_{1}}\hat{u}_{\hat{X}_{2}}\right)_{\hat{X}_{1}} \right] = 0,
\end{align}
where $\hat q \equiv q/L$ and $\hat{X}_1$, $\hat{X}_2$ are defined as in the previous sections.  

Guided by the results of the previous subsection, we know that we must perform the changes of function and variables defined by \eqref{changes2}, if we aim at deriving the Z equation from the last equation in \eqref{equation}. 
Here, those changes give the first terms in the expansion of the two equations \eqref{two} as
\be
\epsilon^2 \hat{q}_{\chi} \pm \alpha \hat q_{\tau} \pm 2C_{01}\epsilon^2 \alpha^2 \left(U^2_{\tau}\right)_{\tau}+\ldots = 0, 
\qquad 
\epsilon \hat{q}_{\eta}+2C_{01}\epsilon^3 \alpha \left(U_{\tau}U_{\eta}\right)_{\tau} + \ldots = 0.
\en 
Now the first of these tells us that $\hat q$ must be of order $\epsilon^2$: $\hat q = \epsilon^2 Q$, say. 
Then the equations \eqref{two} reduce further to
\be \label{comp}
Q_{\tau} + 2C_{01} \alpha \left(\hat{U}^2_{\tau}\right)_{\tau}= 0, \qquad 
Q_{\eta} + 2C_{01} \alpha \left(\hat{U}_{\tau} \hat{U}_{\eta}\right)_{\tau} = 0,
\en 
at order $\epsilon^2$.
Clearly these expressions for the derivatives of $Q$ are not compatible. 

We may thus conclude that linearly-polarized waves are not possible in general incompressible elastic materials. 
Indeed, Horgan and Saccomandi (2003) explain that when an anti-plane shear deformation occurs in a general incompressible material, it is necessarily accompanied by secondary motions associated with in-plane deformations. 
This coupling increases dramatically the complexity of the calculations and is an obstruction to the propagation of linearly polarized two-dimensional shear waves.
In general incompressible materials it is necessary to take into account the in-plane motions coupled to the two-dimensional shear wave.

In fact, Wochner et al. (2008) acknowledge as much when they correct Enflo et al. (2006) and their study of linearly-polarized nonlinear shear waves, and conclude that ``Simply stated, quadratic nonlinearity prevents the existence of a pure linearly polarized shear wave beam.'' 
However, they then state that ``Cubic nonlinearity can be more pronounced than quadratic nonlinearity due to the quasiplanar nature of wavefronts in directional sound beams'', and go on to ``consider only the effects of cubic nonlinearity.''
But ignoring quadratic terms in favour of cubic terms runs contrary to the very idea of asymptotic expansions. 
Upon examination of their equation of motion (11), we see that the only way to make quadratic terms vanish is to take $4\mu+A = 0$. 
According to \eqref{connections} here, this is equivalent to saying that $C_{01}=0$, making $W$ in \eqref{incomp0} belong to the class of generalized neo-Hookean materials.


\section{Dissipation and Dispersion}
\label{dissip-disper}


In our treatment of dissipative and dispersive effects, we use the following constitutive law for the Cauchy stress:
\be \label{second}
\vec{\sigma}=\vec{\sigma}^{E}+\vec{\sigma}^{D},
\en
where 
\be \label{second2}
\vec{\sigma}^{E} = -p\vec{I} + \mu\left[1+\gamma(I_1-3)\right]\vec{B}, 
\qquad
\vec{\sigma}^{D}= \nu \vec{A}_1 + \alpha_1(\vec{A}_2 - \vec{A}_1^2),
\en
which separates the purely elastic part $\vec{\sigma}^E$ of the stress from a viscoelastic part $\vec{\sigma}^D$.

More specifically, $\vec{\sigma}^E$ coincides with the Cauchy stress of an elastic Yeoh material with strain energy \eqref{yeoh}. 
We make this choice for convenience, but it is clear that once the asymptotic expansions are performed below, the Yeoh form recovers the entire class of generalized neo-Hookean solids, through the identifications: 
\be \label{ident}
\mu[1+\gamma(\lambda^2+2\lambda^{-1}-3)] = 2  w^{(1)}, \qquad
\mu \gamma = 2 w^{(2)}.
\en

The viscoelastic part of the stress, $\vec{\sigma}^D$, is the stress of a second-grade viscoelastic solid, where $\nu>0$, $\alpha_1>0$ are constants and $\vec{A}_1$, $\vec{A}_2$ are the first two Rivlin-Ericksen tensors, defined by 
\be
\vec{A}_1 = \vec{L} + \vec{L}^T, \qquad 
\vec{A}_2 = \vec{\dot A}_{1} + \vec{A}_{1}\vec{L} + \vec{L}^T\vec{A}_{1},
\en
with $\vec{L}= \vec{\dot F}\vec{F}^{-1}$, the displacement gradient.
Here the superposed dot represents the material time derivative.
The first part of $\vec{\sigma}^D$ is $\nu \vec{A}_1 = 2 \nu \vec{D}$, where $\vec{D}$ is the stretch tensor; 
that part describes \emph{dissipation} by Newtonian viscosity effects, and coincides with the dissipative term in the Navier-Stokes theory. 
The second part of $\vec{\sigma}^D$ is $\alpha_1(\vec{A}_2 - \vec{A}_1^2)$;
it describes \emph{dispersion} by accounting for the intrinsic characteristic lengths of the solid, see Destrade and Saccomandi (2006) for details. 

Note that the temporal evolution of the  canonical energy $E(t)$ for our solid \eqref{second}-\eqref{second2} is given by $\dot{E}(t) = -2 \int_{\Omega_t} \nu (\vec{D} \cdot \vec{D})\text{d}V <0$, where $\Omega_t$ is a given region occupied by the material at time $t$.  
It is independent of $\alpha_1$ (Fosdick and Yu, 1996).

With respect to two-dimensional anti-plane motion, the derivation of the governing equation is not overly complicated by the additional terms.
For ease of exposition, we consider that the solid is not pre-strained: $\lambda=1$ 
(It follows that once the asymptotic expansions are performed, the identifications \eqref{ident} reduce to $\mu = 2  w^{(1)}$, $\gamma = 2 w^{(2)}/w^{(1)}$.)

We find that the equations $q_\alpha=0$ ($\alpha=1,2$) still apply here, whilst the last equation in  \eqref{eqn} is changed to  
\be \label{equation10}
\left[ \mu u_\beta + \mu \gamma (u_1^2 + u_2^2)u_\beta  + \nu u_{\beta t} + \alpha_1 u_{\beta tt}\right]_\beta = \rho u_{tt}.
\en
In dimensionless form it becomes
\be \label{equation1bis}
\hat{u}_{ii} 
 + \gamma[(\hat{u}_1^2 + \hat{u}_2^2) \hat{u}_i]_ i + \dfrac{\nu}{\mu T}\hat{u}_{ii \hat{t}} + \dfrac{\alpha_1}{\mu T^2}\hat{u}_{ii \hat{t}\hat{t}}  - \dfrac{\rho L^2}{\mu T^2} u_{\hat{t}\hat{t}} =0,
\en 
using the usual scalings.

When the solid is not dissipative ($\nu=0$) and not dispersive ($\alpha_1=0$), the changes of function and variables \eqref{changes2} give a governing equation at order $\epsilon^3$, when $\alpha$ is chosen as $\alpha = \sqrt{\rho L^2/(\mu T^2)}$.
Here the expansions of the dissipative and dispersive terms in \eqref{equation1bis} start as
\begin{align}
& \frac{\nu}{\mu T}\left(\hat{u}_{11 \hat{t}} + \hat{u}_{22\hat{t}}\right) = \epsilon \dfrac{\nu}{\mu T} \left(\dfrac{\rho L^2}{\mu T^2}\right) \hat{U}_{\tau \tau \tau} +\ldots, 
\notag \\
&  \dfrac{\alpha_1}{\mu T^2} \left(\hat{u}_{11 \hat{t}\hat{t}} + \hat{u}_{22 \hat{t}\hat{t}}\right) = \epsilon  \dfrac{\alpha_1}{\mu T^2} \left(\dfrac{\rho L^2}{\mu T^2}\right) \hat{U}_{\tau \tau \tau \tau} + \ldots.
\end{align}
Clearly, in order to be incorporated into the governing equation for the motion, the dissipative and the dispersive effects must be at least of order $\epsilon^2$: $\nu = \epsilon^2 \hat \nu + \ldots$ and  $\alpha_1 = \epsilon^2 \hat{\alpha}_1 + \ldots$ say, where $\hat \nu$ and $\hat{\alpha}_1$ are constants. 
With that assumption, we find the following form of the equation of motion for the velocity $U = \hat{U}_\tau$,
\be \label{equation11}
\left(\pm U_{\chi} + \beta U^2 U_{\tau}  + \beta' U_{\tau \tau} + \beta'' U_{\tau \tau \tau}\right)_\tau + \demi U_{\eta \eta} = 0,
\en
where
\be
\beta \equiv \dfrac{3\gamma}{2} \left(\dfrac{\rho L^2}{\mu T^2}\right)^{\troismi}, \qquad
\beta' \equiv \dfrac{\hat \nu}{2\mu T}\left(\dfrac{\rho L^2}{\mu T^2}\right)^{\demi}, \qquad 
\beta'' \equiv \dfrac{\hat{\alpha}_1}{2\mu T^2} \left(\dfrac{\rho L^2}{\mu T^2}\right)^{\demi}.
\en
This equation covers not only the Z equation (at $\beta' = \beta'' = 0$), but also the dissipative mZK equation (at $\beta'' = 0$) and the dispersive  mKP equation (at $\beta'=0$).  

To the best of our knowledge this the first time that a mKP equation is derived in the continuum mechanics of \emph{homogeneous solids}.



If we consider the diffusive version of \eqref{equation11}, i.e.
\be \label{equation12}
\left(\pm U_{\chi} + \beta U^2 U_{\tau}  + \beta' U_{\tau \tau}\right)_\tau + \demi U_{\eta \eta} = 0,
\en
and we seek solutions which propagate at an angle $\gamma$ with respect to the $\tau$ direction, it is possible to obtain an interesting reduction.  
To this end let us consider solutions which depend only on $\chi$ and $\tilde{\tau}=\tau+\sqrt{2} \eta \tan (\gamma)-\chi \tan^2(\gamma)$ to reduce \eqref{equation12} to 
\be \label{equation13}
\left(\pm U_{\chi} + \beta U^2 U_{\tilde{\tau}}  + \beta' U_{\tilde{\tau} \tilde{\tau}}\right)_{\tilde{\tau}} = 0.
\en
This is the Nariboli-Lin modified diffusive Burgers' equation (Nariboli and Lin, 1973). The plane wave solution of this equation is computed as
$$
U^{-2}=\frac{\beta}{V_0}+\exp\left[-\frac{2V_0}{\beta'}(\tilde{\tau}-V_0 \chi) \right],
$$
where $V_0$ is the speed of the plane wave. 
This solution is quite different from the plane wave solution of the classical Burgers equation which is expressed in terms of the hyperbolic tangent function.

If we consider the purely dispersive case and we rescale some of the variables and $U=\lambda_1 \tilde{U}, \chi=\lambda_2 \tilde{U}$ and $\eta= \lambda_ 3 \tilde{\eta}$ it is always possible to choose the scaling such that   
\be \label{equation14}
\left(\pm \tilde{U}_{\chi} + \tilde{U}^2 \tilde{U}_{\tilde{\tau}} + \tilde{U}_{\tilde{\tau} \tilde{\tau} \tilde{\tau}}\right)_{\tilde{\tau}} + \tilde{U}_{\tilde{\eta} \tilde{\eta}} = 0,
\en
the modified Kadmotsev-Petviashvili equation. A connection between this equation and the Modified Korteweg de Vries equation may be established with a similar transformation to the one used to reduce \eqref{equation12} to \eqref{equation13}.  

It is also possible to seek direct solutions of \eqref{equation14} in the form
$$
\tilde{U}=F(\zeta), \quad \zeta=a \tilde{\tau}+b \tilde{\eta}-\omega\chi.
$$
This ansatz produces the reduction
\be \label{trav}
a^4 F''=-\frac{a^2}{3}U^3+\left(a\omega -b^2\right)U+c_0\zeta+c_1,
\en
where $c_0, \, c_1$ are integration  constants. 

If we consider solitary waves solutions the asymptotic boundary conditions require $c_0=c_1=0$ and
the usual \textit{energy} integral is given by
$$
\frac{1}{2}U'^2=\frac{U^2}{a^2}\left[\frac{a\omega -b^2}{2}-\frac{U^2}{12}\right]+c_2,
$$ 
where again $c_2$ is an integration constant. It is clear that it must be $c_2=0$ and if $a\omega -b^2 > 0$ an explicit solitary wave solution is given by
$$
F(\zeta)=\left\{6\left(a\omega -b^2\right)\text{sech}^2\left[\frac{2\sqrt{a\omega -b^2}}{2a} (\zeta-\zeta_0) \right] \right\}^{1/2},
$$
where $\zeta_0$ is an integration constant.

If we consider in \eqref{trav} $c_0=0$ and $c_1 \neq 0$ it is also possible to find cnoidal wave solutions in terms of elliptic functions.


\section{Conclusions}


We provide a clear and complete derivation of various model equations for shear waves in the framework of nonlinear incompressible elasticity. 
The rationale for investigating this type of problem is dictated by the use of imaging techniques of improved quality, such as latest generation elastography, to evaluate the nonlinear material properties of soft tissues. 

We show that it is possible to derive a pure (scalar) Z equation  only in the special case of generalized neo-Hookean materials. For general incompressible materials the Z equation is coupled with a non-negligeable transverse motion, an important issue which is often overlooked in the literature. 
Clearly, a rigorous evaluation of the validity of the perturbative expansion requires an extension of the existence, uniqueness and stability results that have been provided so far only for model equations with quadratic nonlinearity.   

We point out that general structure of the various equations here obtained is as follows
$$
\dfrac{\text{d}}{\text{d} \chi} \mathcal{E}(U)+c\nabla_{\bot}U=0,
$$
where $c$ is a suitable material parameter, $\nabla_{\bot}$ is the second order derivative (Laplace operator) in an orthogonal direction to the beam axis and $\mathcal{E}(U)=0$ is the equation governing plane wave propagation (inviscid Burgers, Burgers or KdV equation). 
It would be an interesting problem to study the various transformation and group analysis properties of such general equation to find exact and explicit solutions for various model equations following a general and complete scheme.


\section*{Acknowledgements}

This work is supported by a Senior Marie Curie Fellowship awarded by the Seventh Framework Programme of the European Commission to the first author.
This publication is based on work supported in part by Award No. KUK-C1-013-04 , made by King Abdullah University of Science and Technology (KAUST) (AG), and also based in part upon work supported by the National Science Foundation under grant DMS-0907773 (AG)




\end{document}